
 \documentstyle{amsppt} 
 \magnification=1200 
 \vsize7.8truein 
  
 \def\Om{\Omega}

 \def\lam{\lambda}

 \def\pa{\partial}

 \def\ds{\displaystyle} 
 \nologo 
 \baselineskip=18pt 
 \TagsOnRight 
 \NoRunningHeads 
 \topmatter 
 \title 
 Mean-Field Approximation of Quantum Systems and Classical Limit 
 \endtitle 
 \author 
 S. Graffi$^*$, A. Martinez $^*$ and M.Pulvirenti $^+$ 
 \endauthor 

 \affil 
 $^*$ Dipartimento di Matematica, Universit\`a di Bologna,
Italy
\\ (e-mail: graffi\@dm.unibo.it, martinez\@dm.unibo.it)
\\ 
 $^+$ Dipartimento di Matematica, Universit\`a di Roma "La
Sapienza", Italy
\\  (e-mail: pulvirenti\@mat.uniroma.it)
 \endaffil 
 \abstract 
 We  prove that, for a smooth two-body potentials, the 
 quantum mean-field approximation to the  nonlinear 
 Schr\"odinger equation of the Hartree type is stable at the classical
limit  $h \to 0$, yielding the classical Vlasov equation. 
 \endabstract 

 \endtopmatter 
 \heading 
 Introduction 
 \endheading 
 Consider a system of $N$ identical classical particles of unit mass, 
 evolving 
 according to the dynamics generated by the following mean-field 
 Hamiltonian 
 $$ 
 \Cal H =\sum_{i=1}^N \frac 12 v_i^2+\frac 1N \sum_{i<j} \varphi 
 (x_i-x_j). 
 \tag 1 
 $$ 
 If the two-body interaction $\varphi$ is sufficiently smooth, the 
 behavior of the system for large 
 $N$ is well understood (Seee for instance Ref.s [1],[2],[3]). 
 Namely, if the initial positions and velocities of the particles 
 are independently and identically distributed according to the 
 probability density 
 $f=f(x,v)$,  in the limit $N\to 
 \infty$ each 
 particle   is distributed at time $t$ 
 according to the distribution 
 $f=f(x,v,t)$, independently from the others. Here $f(x,v,t)$ is the 
 solution of the Vlasov equation 
 $$ 
 (\pa_t+v\cdot \nabla_x+E\cdot \nabla_v) f(x,v,t)=0, 
 \tag 2 
 $$ 
 where 
 $$ 
 E(x,t)=-\nabla \varphi*\rho (x,t) 
 \tag 3 
 $$ 
 and 
 $$ 
 \rho(x,t)=\int dv f(x,v,t). 
 \tag 4 
 $$ 
 The quantum analogue of this result has been proved in [4]. The 
 framework is a 
 system of $N$ identical bosons interacting by a mean-field potential 
 energy 
 $$ 
 U(x_1, \dots ,x_N)=\frac 1N \sum_{i<j} \varphi (x_i-x_j). 
 \tag 5 
 $$ 
 Then, if the initial wave function factorizes in the limit $N\to \infty 
 $, 
 each 
 particle evolves according to the following nonlinear Schr\"odinger 
 equation 
 of the Hartree type: 
 $$ 
 (i\hbar \pa_t+\frac{\hbar ^2}{2}\Delta -\varphi * \rho) \psi(t)=0 
 \tag 6 
 $$ 
 where 
 $$ 
 \rho(x,t)=|\psi (x,t)|^2. 
 \tag 7 
 $$ 
 and $\ast$ denotes convolution. Further results concerning the 
 Coulomb interaction (see [5], [6]) have been also proved. 

 In all these results, however, the convergence is strongly dependent on 
 $\hbar$. This dependence prevents  (see the case of the Kac 
 potential below) all situations in which $N\to \infty$ entails $\hbar 
 \to 0$ 
   and the system is therefore asymptotically classical. 

 In the present paper we address precisely this problem and show that the 
 Vlasov 
 equation is indeed recovered in the limit $N\to \infty$ even when $\hbar 
 \to 0$ 
 according to an arbitrary law. 

 The problem has been approached and solved in Ref. [7], where 
 the the result is obtained via compactness techniques, under 
 the hypothesis that the Fourier transform of $\varphi$ is 
 compactly supported. Here we deal with $C^2$ potentials and 
 compute the explicit rate of convergence by means of a 
 constructive method. 

 Our technique is based on the WKB method applied to the 
 $N$-particle system. In the same spirit of Ref. [8] where the 
 classical limit for a class of nonlinear Schr\"odinger 
 equation has been investigated, we choose the initial wave function
in such a way that its phase fulfills the  
 classical Hamilton-Jacobi equation. Consequently the equation 
 for the amplitude, which in our context can be  complex 
 valued, satisfies an equation of hydrodynamical type.  For it 
 we deduce $H^s$ estimates via the energy method. The 
 regularity estimates we obtain are the key to prove the 
 convergence of the Wigner transform of the solution in a 
 rather straightforward way.

 \heading 
 1. Problem and results 
 \endheading 

 Consider an $N$- particle quantum system described by the following 
 mean-field 
 Hamiltonian: 
 $$ 
 H_N= -\frac {h^2}2 \sum_{j=1}^N \Delta_j+ \frac 1N \sum_{i<j} \varphi 
 (x_i-x_j), 
 \quad x_k \in \Bbb R^d 
 \tag 1.1 
 $$ 
 where $\displaystyle \Delta_j =\sum _{\alpha=1}^3  \frac {\pa^2}{\pa 
 x_{j,\alpha}^2}$ and 
 $\varphi$ is a two-body smooth potential. 

 We are interested in the asymptotic behavior of the system at the limit 
 when $N\to 
 \infty$ and  $h\to 0$ simultaneously. 
 \remark {Example: the Kac potential} 
 Consider a system of $N$ identical particles of mass $m=1$ interacting 
 through 
 the Kac potential 
 $$ 
 V_\lam (x)=\frac 1{\lam} \varphi \left(\frac x{\lam}\right) 
 \tag 1.2 
 $$ 
 where  $\lam$ is a large parameter of the same order of $N$ and 
 $\varphi$ is a given smooth potential. The hamiltonian is: 
 $$ 
 H_N= -\frac {\hbar ^2}2 \sum_{j=1}^N \Delta_j+  \sum_{i<j} V_\lam 
 (x_i-x_j). 
 \tag 1.3 
 $$ 
 After the rescaling $x=\lam q $ the hamiltonian becomes: 
 $$ 
 H_N= -\frac 12 \left(\frac {\hbar}{\lam}\right)^2 \sum_{j=1}^N \Delta_j+ 
 \lam^{-1} 
 \sum_{i<j} 
 \varphi (q_i-q_j). 
 \tag 1.4 
 $$ 
 Setting $\lam=N$, $\ds h=\frac {\hbar} {\lam}=\frac {\hbar} {N}$ we 
 finally 
 get: 
 $$ 
 H_N= -\frac {h^2}2 \sum_{j=1}^N \Delta_j+ \frac 1N \sum_{i<j} \varphi 
 (q_i-q_j). 
 \tag 1.5 
 $$ 
 \endremark 
 \bigskip 
 \smallskip 
 \smallskip 
 The initial condition $\Psi_N =\psi^{\otimes N}$ is assumed to be 
 factorized 
 (and hence symmetric in the exchange of particles). 

 For the one-particle wave function we choose initially a WKB state: 
 $$ 
 \psi (x)=a(x) e^{i\frac {\sigma(x)}{h}} 
 \tag 1.6 
 $$ 
 where $a$ and $\sigma$ are smooth and independent of 
 $h$. $\sigma$ is real (more generally, one could also consider 
 the case where $a\sim \sum_{j\geq 0}h^ja_j(x)$ is a 
 semiclassical symbol). 

 Setting $\ds W_N=\frac 1N \sum_{i<j} \varphi (x_i-x_j)$ , we denote 
 $\Psi_N(\cdot,t)$ 
 the solution of the Schr\"o\-din\-ger equation 
 $$ 
 (ih \pa_t+\frac{h^2}{2}\Delta _N-W_N) \Psi_N(t)=0 
 \tag 1.7 
 $$ 
 and introduce its Wigner transform: 
 $$ 
 \split 
 f^N(X_N,V_N,t):=\qquad\qquad\qquad\qquad\qquad 
 \\ 
 \left(\frac 1{2\pi}\right)^{3N} 
 \int dY_N e^{-i Y_N \cdot V_N } \Psi_N( X_N+\frac h2 Y_N,t) \bar 
 \Psi_N( X_N-\frac 
 h2 Y_N,t). 
 \endsplit 
 \tag 1.8 
 $$ 
 Hereafter we use the shorthand notation $X_N=(x_1, \dots ,x_N)$, 
 $Y_N=(y_1, \dots 
 ,y_N)$, $V_N=(v_1, \dots ,v_N)$. 

 We also introduce the $j$-particle Wigner functions defined by: 
 $$ 
 f^N_j(X_j,V_j,t)= (1/2\pi)^{3(j-N)}
 \int dX_{N-j} 
 \int dV_{N-j}f^N(X_N,V_N,t)= 
 $$ 
 $$ 
 (\frac 1{2\pi})^{3j} \int dX_{N-j} 
 \int dY_j  e^{-i Y_j \cdot V_j } \Psi_N( X_N+\frac h2 Y_j,t) 
 \bar \Psi_N( X_N-\frac 
 h2 Y_j,t) 
 \tag 1.9 
 $$ 
 where $X_{N-j}=(x_{j+1}, \dots ,x_N), V_{N-j}=(v_{j+1}, \dots ,v_N)$, 
 $Y_j=(y_1, \dots ,y_j)$ and $ X_N+hY_j=(x_1+ hy_1, \dots ,x_j+h y_j, 
 x_{j+1}, \dots, x_N)$. Note that, at time $t=0$, 
 $f^N_j=f_0^{\otimes j}$ where $f_0$ is the Wigner transform of 
 $\psi$. Hence $$ 
 f_0 
 \to\rho (x)\delta(v-u(x)) \quad \hbox {in} \;\;\Cal{D'} \;\; 
 \hbox {as}\;\; 
 h\to 0 
 \tag 1.10 
 $$ 
 where $\rho=\vert a\vert^2$ and $u=\nabla \sigma$ ($\rho=\vert 
 a_0\vert^2$ in the 
 case where $a\sim\sum_{j\geq 0}h^ja_j(x)$). 
 \remark {Remark} We remark that $f^N_j$ is the Wigner transform of the 
 reduced 
 density matrices: 
 $$ 
 \rho (X_j,Z_j) =\int dX_{N-j} \bar \Psi (X_j,X_{N-j}) \Psi 
 (Z_j,X_{N-j}). 
 $$ 
 \endremark 
 \bigskip 
 The asymptotics of $ f^N_j(t) $ is described by the following theorem: 
 \proclaim {Theorem 1.1} 
 Assume $\varphi \in C^{2}(\Bbb R^3)$, $\sigma \in 
 C^{2}(\Bbb 
 R^3)$, $\ds \partial^{\alpha}\varphi$, $\ds \partial^{\alpha}\sigma$ 
 uniformly 
 bounded for $|\alpha|\leq 2$, 
 and 
 $a\in C^{2}\cap H^2 (\Bbb R^3)$.  Let 
 $h=h(N) \to 0$ as $N \to \infty$. Then there exists 
 $T>0$, sufficiently small, such that for all $j\in \Bbb N$: 
 $$ 
 f^N_j(t)\to f_j(t)\qquad \hbox {in } \Cal D '(\Bbb R^{3j} \times \Bbb 
 R^{3j}), 
 \quad t\in [0,T] 
 \tag 1.11 
 $$ 
 where $f_j (t)=f (t)^{\otimes j}$ and $f(t)$ is the unique (weak) 
 solution of the 
 classical Vlasov equation: 
 $$ 
 (\pa_t+v\cdot \nabla_x+E\cdot \nabla_v) f(x,v,t)=0. 
 \tag 1.12 
 $$ 
 Here: 
 $$ 
 E(x,t)=-\nabla \varphi * \rho (x,t),\qquad \rho (x,t)=\int dv f(x,v,t). 
 \tag 1.13 
 $$ 
 Moreover, for $t\in [0,T]$: 
 $$ 
 f(x,v,t)= \rho (x,t) \delta (v-u(x,t)) 
 \tag 1.14 
 $$ 
 where the pair $(\rho, u)$ fulfills the continuity and the momentum 
 balance 
 equations: 
 $$ 
 \pa_t \rho+\hbox {div} (u\rho)=0; \quad \pa_t u+u\cdot \nabla u=-\nabla 
 \phi*\rho. 
\tag 1.15a
 $$ 
 More precisely, for any test function $F\in \Cal D (\Bbb R^{3j} \times 
 \Bbb 
 R^{3j})$, one has 
 $$ 
 \langle f^N_j(t)-f_j(t),F\rangle =\Cal O(h+N^{-1}) 
 \tag 1.15b 
 $$ 
 for $h$ small enough and $N$ large enough. 
 \endproclaim

 We refer to this particular situation as hydrodynamic for obvious 
 reasons. 
 \newline 
 Theorem 1.1 is based on some regularity estimates to be  established in 
 the 
 next section. We will write the solution of the Schr\"o\-din\-ger 
 equation (1.7) under the form 
 $$ 
 \Psi_N(X_N,t)=A^N(X_N,t) e^{i\frac {S^N(X_N,t)}{h}} 
 \tag 1.16 
 $$ 
 where $S^N$ is the classical 
 action 
 satisfying the Hamilton-Jacobi equation: 
 $$ 
 \pa_t S^N +\frac 12 |\nabla S^N|^2 +W_N=0 
 \tag 1.17 
 $$ 
 and, consequently, $A_N$ is the solution of the "transport" equation: 
 $$ 
 \pa_t A^N+(P^N \cdot \nabla )A^N +\frac 12 A^N \hbox {div} P^N= -h 
 \frac i2 \Delta A^N. 
 \tag 1.18 
 $$ 
 Here $P^N:=\nabla S^N$ satisfies: 
 $$ 
 \pa_t P^N+(P^N \cdot \nabla )P^N = -\nabla W. 
 \tag 1.19 
 $$ 
 As we shall show later on, the above representation holds for a short 
 time 
 $T>0$ which may be chosen uniformly in $N$. 

 We then consider initial data which are not of hydrodynamical type. 
 We restrict ourselves to classical data, namely those for which the 
 Wigner 
 transform is positive and normalized: 
 $$ 
 f^N(X_N,V_N)=f^{\otimes N}(X_N,V_N) 
 \tag 1.20 
 $$ 
 where $f$ is a classical probability distribution on the
one-particle phase space. 

 Consider now a one-particle superposition of particular WKB states, 
 given by the following  density matrix: 
 $$ 
 \rho (x,y)=\int dw \quad e^{i\frac wh \cdot (x-y)} a(x;w)\bar 
 a(y,w). \tag 1.21 
 $$ 
 The Wigner transform of  the density matrix is: 
 $$ 
 f_{\rho} (x,v)= (\frac 1{2\pi})^{3} \int dw \int  dz 
 e^{iz\cdot (v-w)} a(x-\frac h2 z;w)\bar a(x+\frac h2 
 z,w)=|a(x,v)|^2 +O(h). \tag 1.22 
 $$ 
 Setting $|a(x,v)| =\sqrt{f(x,v)}$ we see that $f$ and $f_{\rho}$ are 
 asymptotically 
 equivalent (in $\Cal D')$) in the limit $h\to 0$. Therefore we assume 
 as initial condition 
 $$ 
 \rho^N (X_N,Y_N)=\prod_{i=1}^N \rho (x_i,y_i) 
 \tag 1.23 
 $$ 
 for the density matrix (1.21) with $a (x;w)=\sqrt {f(x,w)}$. 
  We will prove 
 \proclaim {Theorem 1.2} 
 Assume $\varphi \in C^{2}(\Bbb R^3)$, $\sigma \in 
 C^{2}(\Bbb 
 R^3)$, $\ds \partial^{\alpha}\varphi$, $\ds \partial^{\alpha}\sigma$ 
 uniformly 
 bounded for $|\alpha|\leq 2$, 
 $a$ compactly supported in $w$ and 
 $a(\cdot,w)\in C^2\cap H^2 (\Bbb R^3)$ for all $w$.   Let 
 $h=h(N) \to 0$ as $N \to \infty$. Then  for all $j\in \Bbb N$ and $t\ge 
 0$: 
 $$ 
 f^N_j(t)\to f_j(t)\qquad \hbox {in } \Cal D '(\Bbb R^{3j} \times \Bbb 
 R^{3j}), 
 \tag 1.24 
 $$ 
 where $f_j (t)=f (t)^{\otimes j}$ and $f(t)$ is the unique 
 solution of the classical Vlasov equation: 
 $$ 
 (\pa_t+v\cdot \nabla_x+E\cdot \nabla_v) f(x,v,t)=0 
 \tag 1.25 
 $$ 
 where 
 $$ 
 E(x,t)=-\nabla \varphi * \rho (x,t),\qquad \rho (x,t)=\int dv f(x,v,t), 
 \tag 1.26 
 $$ 
 with initial datum $f(x,v)=|a(x,v)|^2$. 
 \endproclaim 
 Note that here the convergence result holds globally in time. 

 \heading 
 2. The classical system and its mean-field properties 
 \endheading 

 Consider the associated Hamiltonian system: 
 $$ 
 \dot X_N(t)=V_N (t);\quad 
 \dot V_N (t)_i=-\frac 1N \sum_{j\neq i} \nabla \varphi (x_i(t)-x_j(t)) 
 \tag 2.1 
 $$ 
 where 
 $$ 
 X_N(t)=(x_1(t)\dots x_N(t)), \quad V_N(t)=(v_1(t)\dots v_N(t)). 
 $$ 
 Denote by
 $X_N(t,X_N,V_N)$, $V_N(t)=V_N(t,X_N,V_N)$ the solution of the Cauchy 
 problem 
 with 
 initial conditions $X_N,V_N$. For a given initial datum $(X_N,V_N)$, we 
 consider the empirical distribution, that is a one-particle time 
 depending 
 measure, defined by: 
 $$ 
 \mu^N (dx,dv,t):=\frac 1N \sum _{i=1}^N \delta (x-x_i(t))\delta 
 (v-v_i(t)) dx 
 dv. 
 \tag 2.2 
 $$ 

 The following facts are well known (see e.g. [1],[2],[3]). 

 1) $\mu^N (dx,dv,t)$ is a weak solution of the Vlasov equation (1.12). 
 Namely, for any test function 
 $h=h(x,v)$, setting $\langle \mu^N (t), h\rangle = \int \mu^N (dx,dv,t) 
 h(x,v)$, we have: 
 $$ 
 \frac {d}{dt} \langle \mu^N (t), h\rangle= \langle \mu^N (t), v\cdot 
 \nabla_x h\rangle - 
 \langle \mu^N (t), \nabla_x \varphi *\mu^N (t) \cdot \nabla_v h\rangle. 
 \tag 2.3 
 $$ 
 This follows by a direct computation. 
 \smallskip 
 2) Weak solutions to the Vlasov equation are continuous with respect to 
 the initial datum in the 
 topology of the weak convergence of the measures. 
 \smallskip 
 In particular 1) and 2) imply that if $\mu^N (0)\to f$ 
 weakly, then $\mu^N (t)\to f(t)$weakly, where $f(t)$ is the unique 
 solution to the Vlasov equation with 
 initial datum $f$. If $f$ is sufficiently regular then $f(t)$ hinerits 
 such a regularity and the solution 
 is classical. 
 \smallskip 
 3) Let $f^{N} (X_N,V_N,0) $ be an initial symmetric $N$-particle 
 distribution and 
 let $f^{N} (X_N,V_N,t)=f^{ N}(X_N(X_N,V_N,-t),V_N(X_N,V_N,-t))$ be 
 the solution of the Liouville equation. Define the $j$-particle 
 marginals by: 
 $$ 
 f^{N}_j(X_j,V_j,t):=\int dX_{N-j} \int dV_{N-j}f^{N} (X_N,V_N,t) . 
 \tag 2.4 
 $$ 
 Then, if 
 $$ 
 f^{N}_j \to f^{\otimes j} 
 \tag 2.5 
 $$ 
 in the limit $N \to \infty$ and in the sense of the weak convergence of 
 the 
 measures, where $f=f(x,v)$ is a given $1$-particle initial distribution, 
 then 
 $$ 
 f^{N}_j(t) \to f^{\otimes j}(t) 
 \tag 2.6 
 $$ 
 weakly, where $f(t)$ solves the Vlasov equation with initial condition 
 $f$. Property 
 (2.6) is called propagation of chaos.

 We now specialize the above results to our hydrodynamical case. We 
 suppose that 
 initially: 
 $$ 
 f(x,v)=f(x,v,0)=\rho (x) \delta (v-u(x)) 
 \tag 2.7 
 $$ 
 (in the sequel $u=\nabla \sigma$) and denote: 
 $$ 
 \Phi^t(X_N):=X_N( X_N,P^N(X_N),t) 
 \tag 2.8 
 $$ 
 where $P^N(X_N):=\{ u(x_i)\}_{i=1}^N$. Then, for a given test 
 function $F_j\in \Cal D( \Bbb R^3 \times \Bbb R^3)$, 
 $$ 
 \int f^{N}_j (t)F_j dX_j dV_j= 
 $$ 
 $$ 
 \int dX_N dV_N f^{\otimes N} (X_N,V_N) 
 F_j(X_N^j (X_N,V_N,t),V_N^j (X_N,V_N,t)) =
 $$ 
 (by the Liouville theorem) 
 $$ 
 =\int dX_N \rho ^{\otimes N} (X_N) F_j(\Phi^t (X_N)^j,\dot \Phi^t 
 (X_N)^j) 
 \to 
 $$ 
 $$ 
 \int dX_j dV_j f ^{\otimes j}(X_j,V_j,t) F_j(X_j,V_j) 
 \tag 2.9 
 $$ 
 in the limit $N\to \infty$. Here we are using the notation $X_N^j$ to 
 indicate 
 the vector $(x_1, \dots ,x_j)$ if $X_N=(x_1, \dots ,x_N)$. On the other 
 hand, if 
 $f(t)$ is the solution of the Vlasov equation, for a short time
$t<T$ it has the 
 form 
 $$ 
 f(x,v,t)=\rho (x,t) \delta (v-u(x,t)) 
 \tag 2.10 
 $$ 
 with $\rho$ and $u$ solution of eq.s (1.15) as follows by a direct 
 computation. Indeed if the pair $(\rho,a)$ solves Eq. (1.15), 
 $f(x,v,t)$ given by
 (2.10) is a solution of the Vlasov equation and its uniqueness
entails the assertion.

 Moreover  observe that 
 the estimate: 
 $$ 
 0<C_1\leq |\nabla_{x_i} \Phi^t(X_N)|\leq C_2 <1 
 \tag 2.11 
 $$ 
 which holds for a short time $t<T$, with constants $C_1$ and $C_2$ 
 independent 
 of $N$ due to the mean-field nature of the interaction, 
 allows us to invert the mapping $X_N \to \Phi^t (X_N)$, so that 
 $S^N,P^N,A^N$ 
 exist 
 for such a time interval. The estimate (2.11) easily follows from the 
 analysis 
 developed in Section 4.

 We conclude this section summarizing under the form of a Proposition 
 some  regularity estimates on the classical flow,  established in 
 Section 4 below, which will be used in the 
 convergence proof.

 \proclaim {Proposition 2.1} 
 Assume $\varphi \in C^{2}(\Bbb R^3)$, $\sigma \in 
 C^{2}(\Bbb 
 R^3)$, $\ds \partial^{\alpha}\varphi$, $\ds \partial^{\alpha}\sigma$ 
 uniformly 
 bounded for $|\alpha|\leq 2$, and 
 $a\in C^2\cap H^2  (\Bbb R^3) $. Let $t\in [0,T]$ with $T$ 
 sufficiently 
 small. Then, there exists $C_1>0$ independent of $N$ such that for
all $N$ and all $j\in \{   1,\dots 
 ,N\}$, one has, 
 $$ 
 \| \nabla_{x_j}A^N(t) \|_{L^2} \leq C_1. 
 \tag 2.12 
 $$ 
 Moreover for $\tau \in [0,t]$,  denoting by $x_i^\gamma$, $\gamma=1,2,3$ 
 the components of $x_i$ 
 $$ 
 |\frac {\pa P^N_k(\Phi^{(t-\tau)}(X_N),t)}{\pa x_i^\gamma}| \leq 
 C\left( \frac 1N +\delta_{i,k} \right) 
 \tag 2.13 
 $$ 
 and 
 $$ 
 |\frac {\pa \Phi^{(t-\tau)}(X_N)_k}{\pa x_i^\gamma}| \leq C\left( \frac 
 1N 
 +\delta_{i,k}\right) 
 \tag 2.14 
 $$ 
 where the constant $C$ is independent of $N$, $k$, and $i$. 

 \endproclaim 

 We note that the estimates (2.13) and (2.14) express the weak dependence 
 of 
 the position and momentum of the $k$-th particle with respect to the 
 position of the 
 $i$-th particle at time 
 $0$, as it is expected in a mean-field theory.

 \heading 
 3. Convergence 
 \endheading 

 We are now in position to prove Theorem 1.1. We first observe that, for 
 a time 
 interval for which estimates (2.11), (2.12), (2.13) and (2.14) hold, we 
 have 
 classical solutions of eq.s (1.17), (1.19) and (1.18). Therefore we can 
 express 
 the $j$- particle Wigner function in terms of $A^N, S^N $ and $P^N$. 
 For $ F_j\in \Cal D (\Bbb R^{3j} \times 
 \Bbb R^{3j})$, we have: 
 $$ 
 \int F_j(X_j,V_j)f^N_j(X_j,V_j,t)dX_j dV_j= 
 $$ 
 $$ 
 \left(\frac 1{2\pi}\right)^ {3j} \int 
 dX_N \int dV_j \int dY_je^{-i Y_j\cdot V_j} 
  F_j(X_j,V_j) 
  \Psi_N( X_N+\frac h2 Y_j,t) \bar \Psi_N( X_N-\frac h2 Y_j,t) 
 $$ 
 $$ 
 =\left(\frac 1{2\pi}\right)^{\frac {3}{2} j} 
 \int dX_N  \int dY_j \tilde  F_j(X_j,Y_j) 
 A^N( X_N+\frac h2 Y_j,t) \bar A^N( X_N-\frac h2 Y_j,t) 
 $$ 
 $$ 
 e^{\frac ih [S^N( X_N+\frac h2 Y_j,t)-S^N(X_N-\frac h2 Y_j,t)]}, 
 \tag 3.1 
 $$ 
 where $\tilde  F_j$ is the Fourier transform of $ F_j$ in the second 
 variable. Changing variable $X_N \to X_N-\frac h2 Y_j$ and using the 
 fact that 
 $$ 
 \int dY_j \sup_{X_j} |\tilde F_j (X_j, Y_j)| |Y_j| 
 +\int dY_j \sup_{X_j} |\nabla_{X_j}\tilde F_j(X_j, Y_j)| \leq C j, 
 \tag 3.2 
 $$ 
 we obtain that (setting $X_j=X_N^j$): 
 $$ 
 (3.1)=(\frac 1{2\pi})^{\frac {3j}{2}} 
  \int dX_N  \int dY_j \tilde F_j(X_j,Y_j) 
 A^N( X_N+hY_j,t) \bar A^N( X_N,t) 
 $$ 
 $$ 
 e^{\frac ih [S^N( X_N+hY_j,t)-S^N(X_N,t)]}+O(h). 
 \tag 3.3 
 $$ 
 Now Lagrange's theorem yields: 
 $$ 
 A^N( X_N+hY_j,t)=A^N( X_N,t)+\int_0^h d\lam \nabla_{X_j} A^N( X_N+\lam 
 Y_j,t)\cdot 
 Y_j 
 $$ 
 Moreover, since 
 $$ 
 \|\nabla_{X_j} 
 A^N\|^2_{L_2}=\sum_{i=1}^j \|\nabla_{x_i} A^N\|^2_{L_2}\leq Cj; \quad 
 \| A^N (t) \|_{L_2}=\| A^N (0) \|_{L_2}=1 
 $$ 
 by (3.2) and Proposition 2.1 we get the estimate 
 $$ 
 \int _0^h d\lam \int dX_N  \int dY_j |\tilde F_j(X_j,Y_j)| 
 |\nabla_{X_j} A^N( X_N+\lam Y_j,t)| | A^N( X_N,t)|\leq 
 $$ 
 $$ 
 h \|A^N\|_{L_2}\|\nabla_{X_j} A^N\|_{L_2} 
 \int dY_j \sup_{X_j} |\tilde F_j (X_j, Y_j)| |Y_j| 
 \leq Ch \sqrt {j}. 
 \tag 3.4 
 $$ 
 Hence we can conclude 
 that: 
 $$ 
 (3.1)= \left(\frac1{2\pi}\right)^{\frac{3j}{2}}\int dX_N\int 
 dY_j\tilde F_j(X_j,Y_j)|A^N( X_N,t)|^2 
 $$ 
 $$ 
 e^{\frac ih [S^N( X_N+ h Y_j,t)-S^N(X_N,t)]} 
 +O( h). 
 \tag 3.5 
 $$ 

 Note that $O(h)$ (as well $O(\frac 1N)$ later on) 
 depends on $j$ which however is fixed. 
 Furthermore: 
 $$ 
 S^N( X_N+ h Y_j,t)-S^N(X_N,t)= 
 \int_0^h d\lam P^N( X_N+\lam Y_j,t)^j\cdot Y_j= 
 $$ 
 $$ 
 hP^N( X_N,t)^j\cdot Y_j+O(h^2) 
 \tag 3.6 
 $$ 
  again by Proposition 2.1. 

  Here $P^N(X_N)^j$ denotes the projection on the $j$-particle subspace 
 of the 
 vector 
 $P^N(X_N)$. 

  Hence 
 $$ 
 (3.1)=  \left(\frac 1{2\pi}\right)^{\frac {3j}{2}} \int dX_N  \int 
 dY_j \tilde F_j(X_j,Y_j) 
 | A^N( X_N,t)|^2 
 e^{-iP^N( X_N,t)^j\cdot Y_j} 
 +O( h)= 
 $$ 
 $$ 
 \int dX_N   F(X_j,P^N(X_N,t)^j)  | A^N( X_N,t)|^2 
 +O( h). 
 \tag 3.7 
 $$

 Setting $\Gamma^N=|A^N|^2$ we have by (1.18): 
 $$ 
 \pa_t \Gamma^N +\hbox {div }(P^N \Gamma^N )=B^N 
 \tag 3.8 
 $$ 
 where 
 $$ 
 B^N=\frac i2 h (\bar A^N \Delta A^N -  A^N \Delta \bar A^N). 
 \tag 3.9 
 $$ 

 The solution of eq. (3.8) has the representation: 
 $$ 
 \Gamma^N (X_N,t)= \rho ^{\otimes N} (\Phi^{-t}(X_N)) J_N (X_N,t) 
 + \int_0^t ds B^N (\Phi^{-(t-s)}(X_N)) J_N (X_N,t-s), 
 \tag 3.10 
 $$ 
 where 
 $$ 
 J_N (X_N,t)=\det |\frac {\pa \Phi^{-t}(X_N))}{\pa X_N}|. 
 \tag 3.11 
 $$ 
 Therefore 
 $$ 
 \int dX_N   F_j(X_j,P^N(X_N,t)^j)  \Gamma^N( X_N,t) = 
 \int dX_N  \rho ^{\otimes N}(X_N) 
 F_j(\Phi^{t}(X_N)^j,\dot \Phi^{t}(X_N)^j) 
 $$ 
 $$ 
 +\int_0^t ds \int dX_N B^N (\Phi^{-(t-s)}(X_N),s) J(X_N,t-s) 
 F_j(X_j,P^N(X_N,t)^j). 
 \tag 3.12 
 $$ 

 The last term in the r.h.s. of (3.12) can be rewritten as: 
 $$ 
 \int_0^t ds \int dX_N B^N (X_N,s) 
 F_j(\Phi^{(t-s)}(X_N)^j,P^N(\Phi^{(t-s)}(X_N)^j,t)) = 
 $$ 
 $$ 
 \frac {ih}{2} \int_0^t ds \int dX_N  (\bar A^N \Delta A^N -  A^N \Delta 
 \bar 
 A^N)(X_N,s) 
 $$ 
 $$ 
 F_j(\Phi^{(t-s)}(X_N)^j,P^N(\Phi^{(t-s)}(X_N)^j,t)) 
 $$ 
 $$ 
 =\frac {ih}{2}\int_0^t ds \int dX_N  (\bar A^N \nabla A^N \cdot \nabla 
 F_j 
 (\dots) 
 - A^N \nabla \bar A^N \cdot \nabla F_j (\dots)), 
 \tag 3.13 
 $$ 
 here we have integrated by parts and made use of a crucial cancellation. 

 We now observe that 
 $$ 
 \sum_{i=1}^N | \nabla_{x_i} 
 F_j (\Phi^{(t-s)}(X_N)^j,P^N(\Phi^{(t-s)}(X_N)^j,t))| 
 \tag 3.14 
 $$ 
 can be bounded by a constant dependent on $j$ but not on $N$. Indeed: 
 $$ 
 \pa_{x_i^\alpha}F_j (\Phi^{(t-s)}(X_N)^j,P^N(\Phi^{(t-s)}(X_N)^j,t))= 
 $$ 
 $$ 
 \sum_{k=1}^j[\nabla_{y_k} 
 F_j(Y_j,P^N(\Phi^{(t-s)}(X_N)^jj,t))|_{Y_j=\Phi^{(t-s)}(X_N)^j}\cdot 
 \frac {\pa \Phi^{(t-s)}(X_N)_k}{\pa x_i^\alpha}+ 
 $$ 
 $$ 
 \nabla_{v_k} 
 F_j(\Phi^{(t-s)}(X_N)_j,V_j,t))|_{V_j=P^N(\Phi^{(t-s)}(X_N)^j)}\cdot 
 \frac {\pa P^N_k(\Phi^{(t-s)}(X_N),t)}{\pa x_i^\alpha}. 
 $$ 
 Then by Proposition 1.1 we have that: 
 $$ 
 \pa_{x^{\alpha}_i} 
 F_j(\Phi^{(t-s)}(X_N)_j,P^N(\Phi^{(t-s)}(X_N)^j,t))=O(1) 
 $$ 
 if $i=1\dots j$, while 
 $$ 
 \pa_{x^{\alpha}_i} F 
 (\Phi^{(t-s)}(X_N)_j,P^N(\Phi^{(t-s)}(X_N)_j,t))=O(\frac 
 1N) 
 $$ 
 if $i>j$. 
 Hence, by (3.14): 
 $$ 
 |\int dX_N  \bar A^N \nabla A^N \cdot \nabla F_j 
 (\dots)|\leq \|A \|_{L_2} 
 \sum_{i=1}^N | \nabla_{x_i} 
 F_j (\dots)| \|\nabla_{x_i}A \|_{L_2}\leq Cj. 
 $$ 
 Therefore 
 $$ 
 (3.1)= 
 \int dX_N  \rho ^{\otimes N}(X_N) 
 F_j(\Phi^{t}(X_N)_j,\dot \Phi^{t}(X_N)_j)+O(h)+O(\frac 1N). 
 \tag 3.15 
 $$ 
 Notice that the first term in the r.h.s. of (3.15) is purely classical 
 so that 
 we can apply the convergence result (2.9) to conclude the proof. 

 The proof of Theorem 1.2 follows along the same lines. 
 Proceeding as as above the Wigner function is in this case: 
 $$ 
 f^N (X_N,V_N) = (\frac 1{2\pi})^{3} \int d\Om_N \int dY_N 
 e^{-iY_N \cdot V_N} 
 $$ 
 $$ 
 A^N(X_N-\frac h2 Y_N ,\Om_N,t )\bar A^N(X_N+\frac h2 Y_N ,\Om_N,t ) 
 $$ 
 $$ 
 e^{\frac ih [S^N(X_N-\frac h2 Y_N ,\Om_N,t )-S^N((X_N+\frac h2 Y_N 
 ,\Om_N,t) ]} 
 \tag 3.16 
 $$ 
 where $A^N$ and $S^N$ are the amplitude and the action parametrized by 
 the 
 initial 
 momenta $\Om_N$. $S^N$ and $A^N$ are the solution of eq.s (1.17) and 
 (1.18) 
 with 
 initial conditions $ S^N (X_N,\Om_N)=\Om_N\cdot X_N$ and $A^N(X_N, 
 \Om_N)=a^{\otimes N }(X_N, 
 \Om_N)$. Therefore for a test function $F_j$ we have: 
 $$ 
 \int F_j(X_j,V_j)f^N_j(X_j,V_j,t)dX_j dV_j= 
 $$ 
 $$ 
 \left(\frac 1{2\pi}\right)^{\frac {3}{2} j} 
 \int d\Om _N \int dX_N  \int dY_j \tilde  F_j(X_j,Y_j) 
 $$ 
 $$ 
 A^N( X_N+\frac h2 Y_j,\Om_N, t) \bar A^N( X_N-\frac h2 Y_j,\Om_N,t) 
 e^{\frac ih [S^N( X_N+\frac h2 Y_j,\Om_N,t)-S^N(X_N-\frac h2 
 Y_j,\Om_N,t)]}. 
 \tag 3.17 
 $$ 

 Proceeding as in the proof of Theorem 1.1 we find: 
 $$ 
 (f^N_j,F)= \int d\Om_N \int dX_N f^{\otimes N}(X_N,V_N) 
 F(X_N(t,X_N,\Om_N), \dot X_N(t,X_N,\Om_N)) 
 $$ 
 $$ 
 +O( h)+O(\frac 1N). 
 \tag 3.18 
 $$ 
 Note that  that $a$ is  assumed compactly supported in $w$ to avoid 
 complications with the integral in the initial momenta. However a 
 sufficiently 
 rapid decay 
 of $ a(\cdot, w) $,  for large $w$, would yield the same result. 
 Of course, once more, everything holds for a small time interval. 
 However now the smallness of the time 
 interval depends only on the smoothness of the potential 
 $\varphi$ 
 (because of the particular form $\sigma (x)=w\cdot x$). Hence  the 
 convergence 
 at time $T$ allows us to extend the argument up to $2T$, using fact
2) of Section 2 and that $f(x,v,T)$ is still compactly  
supported in   velocity. Therefore the convergence can be extended
to arbitrary times   and the 
 proof of Theorem 1.2 is now completed.

 \heading 
 4. Regularity estimates 
 \endheading 

 In this section we prove Proposition 2.1. We start by considering 

 $$ 
 \pa_t P^N+(P^N \cdot \nabla )P^N = -\nabla W. 
 \tag 4.1 
 $$ 
 which is independent of 
 $$ 
 \pa_t A^N+(P^N \cdot \nabla )A^N +\frac 12 A^N \hbox {div} P^N= -h 
 \frac i2 \Delta A^N. 
 \tag 4.2 
 $$ 
 to be considered later on. 

 Notice that, for $s\in [0,t]$, 
 $$ 
 P_i^N(\phi^{(t-s)}(X_N),t)=\dot \Phi^t(Y_N(s))_i, \quad \hbox 
 {where}\quad 
 Y_N(s)=\Phi^{-s}(X_N),\quad i=1\dots N. 
 \tag 4.3 
 $$ 

 Using now the short-hand notation $x_i(t)=\Phi^t(X_N)_i$, 
 $p_i(t)=\dot \Phi^t(X_N)_i$ and denoting  $x_i^\alpha (t)$ and 
 $p_i^\alpha 
 (t)$ 
 the 
 $\alpha$-th components, $\alpha=1,2,3$, we have: 
 $$ 
 x_i(t)=x_i+\int_0^t  p_i(s) ds 
 $$ 
 $$ 
 p_i(t)=\nabla \sigma (x_i)-\int_0^t  \frac 1N \sum_{k\neq i} \nabla 
 \varphi 
 (x_i(s)-x_k(s)) ds. 
 \tag 4.4 
 $$ 
 where $X_N=(x_1\dots x_N)$. 

 Introducing the force $F^\alpha =-\pa_{x^\alpha}\varphi$, we have: 
 $$ 
 \frac {\pa x_i^\beta (t)}{\pa 
 x_j^\gamma}=\delta_{i,j}\delta_{\beta,\gamma}+\int_0^t ds 
 \frac {\pa p_i^\beta (s)}{\pa x_j^\gamma} 
 \tag 4.5 
 $$ 
 $$ 
 \frac {\pa p_i^\beta (t)}{\pa x_j^\gamma}=\frac {\pa^2 \sigma}{\pa 
 x_j^\gamma \pa x_i^\beta} \delta_{i,j}+\int_0^t  ds 
  \frac 1N \sum_{k\neq i} \sum_\alpha \pa_{x^\alpha} F^\beta 
 (x_i(s)-x_k(s)) 
 \left(\frac {\pa x_i^\alpha (s)}{\pa x_j^\gamma}- 
 \frac {\pa x_k^\alpha (s)}{\pa x_j^\gamma}\right). 
 \tag 4.6 
 $$ 
 Hence: 
 $$ 
 \frac {\pa p_i^\beta (t)}{\pa x_j^\gamma}=\frac {\pa^2 \sigma}{\pa 
 x_j^\gamma \pa x_i^\beta} \delta_{i,j}+ 
 \int_0^t  ds 
 \frac 1N \sum_{k\neq i} \sum_\alpha \pa_{x^\alpha} F^\beta 
 (x_i(s)-x_k(s)) 
 (\delta_{i,j}-\delta_{k,j})\delta_{\alpha,\gamma}+ 
 $$ 
 $$ 
 \int_0^t  ds \int _0^s d\tau 
 \frac 1N \sum_{k\neq i} \sum_\alpha \pa_{x^\alpha} F^\beta 
 (x_i(s)-x_k(s)) 
 \left(\frac {\pa p_i^\alpha (\tau)}{\pa x_j^\beta}-\frac {\pa p_k^\alpha 
 (t)}{\pa 
 x_j^\beta}\right). 
 \tag 4.7 
 $$ 
 We now observe that, if $i\neq j$, the first two terms in the r.h.s. of 
 (4.7) are 
 $O(\frac1N)$ and hence (taking $t\in [0,T]$ with $T$ small enough): 
 $$ 
 \left|\frac {\pa x_i^\beta (t)}{\pa x_j^\gamma}\right|+\left 
 |\frac {\pa p_i^\beta (t)}{\pa 
 x_j^\gamma}\right|\leq C\left(\frac 1N+\delta_{i,j}\right), 
 \tag 4.8 
 $$ 
 with $C$ independent of $N$. 

 \remark {Remark} 
 Higher derivatives could be handled in the same way to obtain: 
 $$ 
 \left|\frac {\pa^s x_i^\beta (t)}{\pa x_{j_1}^{\gamma_1}\dots \pa 
 x_{j_s}^{\gamma_s} 
 }\right|+ 
 \left|\frac {\pa^s p_i^\beta (t)}{\pa x_{j_1}^{\gamma_1}\dots \pa 
 x_{j_s}^{\gamma_s} 
 }\right| 
 \leq 
 C\left(\frac 1N+\prod _{r=1}^s\delta_{i,j_r}\right), 
 \tag 4.9 
 $$ 
 assuming a stronger regularity. 
 \endremark 

 Furthermore, setting $Y_N(s)=(y_1(s), \dots,y_N(s))$, 
 $$ 
 \frac {\pa P_i^\alpha (Y_N,t)}{\pa x_{j}^{\gamma}}= \sum_k \sum_{\beta} 
 \frac {\pa \dot \Phi^t (Y_N)^{\alpha}_i}{\pa 
 y_{k}^{\beta}}|_{Y_N=Y_N(s)} 
 \frac {\pa y_k^{\beta} (s)}{\pa x_j^\gamma}. 
 \tag 4.10 
 $$ 

 We note that  the terms in the sum with $k\neq i$ and $k\neq j$ are 
 $\ds O(\frac 1{N^2})$, while the term with $k=i$ or $k=j$ are $\ds 
 O(\frac 
 1{N})$. Therefore, if $i\neq j$, the full sum is $\ds O(\frac 1{N})$. If 
 $i=j$ 
 the sum is 
 $O(1)$ because of the term $k=i=j$ which is indeed $O(1)$. Summarizing: 
 $$ 
 \left|\frac {\pa P_i^\alpha (\Phi ^{(t-s)}(X_N),t)}{\pa 
 x_{j}^{\gamma}}\right| \leq C\left(\frac 
 1N+\delta_{i,j}\right). 
 \tag 4.11 
 $$ 

 \remark {Remark} 
 A similar analysis on the higher derivatives yields: 
 $$ 
 \left|\frac {\pa^s P_i^\beta (\Phi ^{(t-s)}(X_N),t)}{\pa 
 x_{j_1}^{\gamma_1}\dots \pa 
 x_{j_s}^{\gamma_s} }\right| 
 \leq C\left(\frac 1N+\prod _{r=1}^s\delta_{i,j_r}\right). 
 \tag 4.12 
 $$ 
 \endremark 

 We now proceed to analyze eq. (4.2) to obtain estimate of the solution 
 in 
 $H^1$. 
 Applying the operator $\nabla_{x_j}$ to the equation, we obtain: 
 $$ 
 \pa_t \nabla_{x_j} A^N+ 
 ( \nabla_{x_j} P^N \cdot \nabla )  A^N +(P^N \cdot \nabla )\nabla_{x_j} 
 A^N +\frac 12 
 \nabla_{x_j} A^N 
 \hbox {div} P^N= 
 $$ 
 $$ 
 - \frac 12 
  A^N 
 \hbox {div} \nabla_{x_j}P^N -h 
 \frac i2 \Delta \nabla_{x_j}  A^N. 
 \tag 4.14 
 $$ 
 In computing 
 $$ 
 \frac {d} {dt} (\nabla_{x_j} A^N,\nabla_{x_j} A^N)= 
 (\pa_t \nabla_{x_j} A^N,\nabla_{x_j} A^N)+(\nabla_{x_j} A^N,\pa_t 
 \nabla_{x_j} A^N) 
 \tag 4.15 
 $$ 
 we realize that, due to the symmetry of $\Delta$, the last term does not 
 give 
 any 
 contribution. 
 Also, the sum of the terms non involving $\nabla_{x_j}P^N$ vanishes: 
 $$ 
 (\nabla_{x_j} A^N,  P^N \cdot \nabla \nabla_{x_j} A^N)+ 
 (P^N \cdot \nabla \nabla_{x_j} A^N,  \nabla_{x_j} A^N)+ 
 $$ 
 $$ 
 \frac 12 (\nabla_{x_j} A^N, \nabla_{x_j} A^N \hbox {div} P^N)+ 
 \frac 12 (\hbox {div} P^N \nabla_{x_j} A^N, \nabla_{x_j} A^N )=0. 
 \tag 4.16 
 $$ 
 Here we use the reality of $P^N$ and the identity: 
 $$ 
 \int  P^N \cdot \nabla | \nabla_{x_j} A^N|^2=-\int \hbox {div} P^N | 
 \nabla_{x_j} 
 A^N|^2 
 \tag 4.17 
 $$ 

 We finally observe that, by eq. (4.11), 
 $$ 
 ( \nabla_{x_j} P^N \cdot \nabla )  A^N= 
 (\nabla_{x_j} P^N_j \cdot \nabla_{x_j} ) A^N + \Cal O 
 (N^{-1}\sum_{k\not= j}\| 
 \nabla_{x_k}A^N\|_{L_2}) 
  \tag 4.18 
 $$ 
 and thus, denoting $\|A^N\|_1 := (\sum_k \| \nabla_{x_k} 
 A^N\|_{L_2}^2)^{1/2}$ (so that 
 $\sum_{k\not= j}\| 
 \nabla_{x_k}A^N\|_{L_2}\leq \sqrt N \|A^N\|_1$), 
 we arrive to the inequality: 
 $$ 
 \frac {d}{dt}  \| \nabla_{x_j}A^N(t)\|_{L_2}^2\leq C \{\| 
 \nabla_{x_j}A^N(t)\|_{L_2}^2 
 + N^{-1/2}\| A^N\|_1\|\nabla_{x_j}A^N(t)\|_{L_2}\} 
 \tag 4.19 
 $$ 
 with $C$ independent of $N$ and $j$. In particular, taking the sum over 
 all $j$, 
 we obtain, 
 $$ 
 \frac {d}{dt}  \| A^N(t)\|_1^2\leq 2C \| A^N(t)\|_1^2. 
 \tag 4.20 
 $$ 
 Since at time zero 
 $\| A^N\|_1^2$ is $\Cal O (N)$ the same conclusion holds on any 
 time interval. Going back to Eq. (4.19), we obtain 
 $$ 
 \frac {d}{dt}  \| \nabla_{x_j}A^N(t)\|_{L_2}^2\leq C \| 
 \nabla_{x_j}A^N(t)\|_{L_2}^2 
 \tag 4.21 
 $$ 
 with a new constant $C$ independent of $N$ and $j$. 
 Then the result follow by observing that
$\|\nabla_{x_j}A^N(0)\|_{L_2}$   is uniformly 
 bounded.

 \end